\documentstyle[aps,amssymb, manuscript]{revtex}
\tightenlines \draft

\begin{document}
\title{Space Time Matter inflation}
\author{
$^{1,2}$ Mariano Anabitarte\footnote{
E-mail address: anabitar@mdp.edu.ar}
and $^{1,2}$Mauricio Bellini\footnote{
E-mail address: mbellini@mdp.edu.ar}}
\address{
$^1$Departamento de F\'{\i}sica,
Facultad de Ciencias Exactas y Naturales,
Universidad Nacional de Mar del Plata,
Funes 3350, (7600) Mar del Plata, Argentina.\\
$^2$ Consejo Nacional de Ciencia y Tecnolog\'{\i}a (CONICET).}

\vskip .5cm
\maketitle
\begin{abstract}
We study a model of power-law inflationary inflation
using the Space-Time-Matter (STM) theory of gravity for
a five dimensional (5D) canonical metric that describes
an apparent vacuum. In this approach the expansion is
governed by a single scalar (neutral) quantum field.
In particular, we study the case where the power of expansion
of the universe is $p \gg 1$. This kind of model is more
successful than others in accounting for galaxy formation.

\end{abstract}

\section{Introduction}

A standard mechanism for galaxy formation is the amplification of primordial
fluctuations by the evolutionary dynamics of spacetime.
The inflationary cosmology is based on the dynamics of a quantum field
undergoing a phase transition\cite{libro}.
The exponential expansion of the scale
parameter naturally
gives a scale-invariant spectrum on cosmological scales, in agreement
with experimental data.
This is one of the many
attractive features of the inflationary universe, particulary in regard
to the galaxy formation problem\cite{2} and it
arises from the fluctuations of the
inflaton, the quantum field which induces inflation.
This field can be semiclassically expanded in terms of its
expectation value plus other field, which describes the quantum
fluctuations\cite{habib}.
The quantum to classical transition of quantum fluctuations has been
studied in thoroughly\cite{4}.
The infrared matter field fluctuations are
classical and can be described by a coarse-grained field which takes
into account only wavelengths larger than the Hubble radius. The dynamics
of this coarse-grained field is described by a second order stochastic
equation, which can be treated using the Fokker-Planck
formalism.
This issue has been the subject of intense work during the last two decades
\cite{5}. Because of the sucess of this theory to explain the large-scale
structure formation, inflation has nowaday become a standard ingredient
for the description of the early universe. In fact, it is the
unique that solves some of the problems of the standard big-bang
scenario and also makes predictions about Cosmic Microwave Background
(CMBR) anisotropies, which are being measured with increasing precision.

On the other hand, recently, extra dimensional theories of gravity have
received much interest, mainly sparked by works in string\cite{a}
and supergravity theories\cite{b}. For the most part, four-dimensional (4D)
space-time has been extended by the addition of several extra spatial dimensions,
usually taken to be compact.
Other very interesting approach, developed by Wesson and co-workers\cite{c}
have given new impetus to the study of 5D gravity.
None of the standard dimensional reduction techniques imposed to
reduce the number of space-time dimensions to four, are
adhered to in their approach; indeed, the extra spatial dimension
is not necessarily assumed to be
compact. The main question they address in whether the 4D properties
of matter can be viewed as being purely geometrical in origin. This idea
is not new, and was originally introduced
by Einstein\cite{d}.

In this work we are interested in studying the early inflationary
dynamics of the universe from the STM theory of gravity.
In particular, we are aimed to study power-law inflation, where
the scale factor of the universe growth as $a \sim t^p$, being
$p>1$ the power of expansion during inflation.
To do this, we use the Ponce de Leon metric\cite{pdl}, in the
limital case where the power of expansion is $p \gg 1$, which
describe an asymptotic de Sitter expansion of the universe.

\section{Basic STM equations}

Following the idea suggested by Wesson and co-workers\cite{c,c1}, in this
section we develope
the induced 4D equation of state from the
5D vacuum field equations, $G_{AB}=0$ ($A,B=0,1,2,3,4$),
which give the 4D Einstein equations $G_{\mu\nu}=
8\pi G \  T_{\mu\nu}$ ($\mu,\nu =0,1,2,3$).
In particular, we consider a 5D spatially isotropic and 3D flat
spherically-symmetric
line element
\begin{equation}\label{1}
dS^2 =  e^{\alpha(\psi,t)} dt^2 - e^{\beta(\psi,t)} dr^2
- e^{\gamma(\psi,t)}
d\psi^2,
\end{equation}
where $dr^2 = dx^2+dy^2+dz^2$ and $\psi$ is the fifth coordinate.
We assume that $e^{\alpha}$, $e^{\beta}$ and $e^{\gamma}$
are separable functions of the variables $\psi$ and $t$.
The equations for the relevant Einstein
tensor elements are
\begin{eqnarray}
G^0_{ \  0}& =& -e^{-\alpha} \left[\frac{3 \dot\beta^2}{4} +
\frac{3\dot\beta \dot\gamma}{4}\right] + e^{-\gamma}\left[
\frac{3 \stackrel{\star\star}{\beta}}{2}
+\frac{3\stackrel{\star}{\beta}^2}{2}-\frac{3\stackrel{\star}{\gamma}
\stackrel{\star}{\beta}}{4}\right],\\
G^0_{ \  4} & = & e^{-\alpha}\left[\frac{3\stackrel{\star\cdot}{\beta}}{2}
+\frac{3\dot\beta
\stackrel{\star}{\beta}}{4}
-\frac{3\dot\beta \stackrel{\star}{\alpha}}{4} - \frac{3
\stackrel{\star}{\gamma} \dot\gamma}{4}\right],\\
G^i_{ \  i} & = &- e^{-\alpha} \left[\ddot\beta + \frac{3\dot\beta^2}{4}+
\frac{\ddot\gamma}{2} + \frac{\dot\gamma^2}{4} + \frac{\dot\beta\dot\gamma}{2}-
\frac{\dot\alpha\dot\beta}{2} - \frac{\dot\alpha\dot\gamma}{4}\right] \nonumber \\
& + & e^{-\gamma}\left[\stackrel{\star\star}{\beta}
+\frac{3\stackrel{\star}{\beta}^2}{4}
+ \frac{\stackrel{\star\star}{\alpha}}{2} + \frac{
\stackrel{\star}{\alpha}^2}{4} +
\frac{\stackrel{\star}{\beta}\stackrel{\star}{\alpha}}{2}
-
\frac{\stackrel{\star}{\gamma}\stackrel{\star}{\beta}}{2}
- \frac{\stackrel{\star}{\alpha}\stackrel{\star}{\gamma}}{4}\right],\\
G^4_{ \  4} & = & e^{-\alpha} \left[\frac{3\ddot\beta}{2} +
\frac{3\dot\beta^2}{2}-
\frac{3 \dot\alpha \dot\beta}{4}\right]
+e^{-\gamma}\left[\frac{3\stackrel{\star}{\beta}^2}{4} +
\frac{3\stackrel{\star}{\beta}\stackrel{\star}{\alpha}}{4} \right],
\end{eqnarray}
where the overstar and the overdot denote respectively ${\partial \over
\partial\psi} $ and ${\partial \over \partial t}$, and
$i=1,2,3$. We shall use the signature $(+,-,-,-)$ for the
4D metric, such that we define
$T^0_{ \  0} = \rho_t$ and $T^1_{ \  1} = -{\rm p}$, where
$\rho_t$ is the total energy density and ${\rm p}$ is the pressure.
The 5D-vacuum conditions ($G^A_B =0$) are given by\cite{WE}
\begin{eqnarray}
&& 8\pi G \rho_t = \frac{3}{4} e^{-\alpha} \dot\beta^2, \label{6} \\
&& 8\pi G {\rm p} = e^{-\alpha} \left[\frac{\dot\alpha\dot\beta}{2} -
\ddot\beta - \frac{3\dot\beta^2}{4}\right], \label{7} \\
&& e^{\alpha} \left[\frac{3\stackrel{\star}{\beta}^2}{4} + \frac{3
\stackrel{\star}{\beta}\stackrel{\star}{\alpha}}{4}\right]=
e^{\gamma} \left[\frac{\ddot\beta}{2} + \frac{3\dot\beta^2}{2} - \frac{
\dot\alpha\dot\beta}{4}\right]. \label{8}
\end{eqnarray}
Hence, from eqs. (\ref{6}) and (\ref{7}) and taking
$\dot\alpha = 0$, we obtain
the equation of state for the induced matter
\begin{equation}\label{9a}
{\rm p} = - \left(\frac{4}{3} \frac{\ddot\beta}{\dot\beta^2} + 1\right)
\rho_t.
\end{equation}
Notice that for $\ddot\beta/\dot\beta^2 \le 0$
and $\left|\ddot\beta/\dot\beta^2
\right| \ll 1$ (or zero), this equation describes an inflationary universe.
The particular case $\ddot\beta/\dot\beta^2 =0$ corresponds to a 4D de Sitter
expansion for the universe.

As in a previous paper\cite{NPB03},
we shall consider power-law inflation, which can be obtained
from the metric
(\ref{1}), when $\alpha$, $\beta$ and $\gamma$ are functions
ob the following coordinates:
\begin{equation}\label{8a}
\alpha \equiv \alpha(\psi); \quad \beta \equiv \beta(\psi,t);
\quad \gamma \equiv \gamma(t).
\end{equation}
Here, $e^{\beta}$ is a separable function of $\psi$ and $t$. The conditions
(\ref{8a}) imply that $\dot\alpha = \stackrel{\star}{\gamma}=0$. Furthermore,
we shall consider the case where
all the coordinates are independent. The choice (\ref{8a})
implies that only the spatial sphere and the the fifth coordinate have
squared sizes $e^{\beta}$ and $e^{\gamma}$, respectively,
that evolve with time.

\section{The model}

In this paper we shall consider the
particular case of choosing for the metric (\ref{1}):
\begin{equation}
e^{\alpha} = \psi^2, \quad e^{\beta}= \left(\frac{t}{t_0}\right)^{2p}
\psi^{\frac{2p}{p-1}}, \quad e^{\gamma} = \frac{t^2}{(p-1)^2},
\end{equation}
which corresponds to the Ponce de Leon metric\cite{pdl}
\begin{equation}\label{pdl}
dS^2 = \psi^2 dt^2 - \left(\frac{t}{t_0}\right)^{2p} \psi^{\frac{2p}{p-1}}
dr^2 - \frac{t^2}{(p-1)^2} d\psi^2,
\end{equation}
for which the absolute value for the
determinant of the metric tensor $g_{AB}$ is
\begin{displaymath}
\left|^{(5)}g\right|
= \left[\frac{t^{3p+1} \psi^{\frac{4p-1}{p-1}}}{(p-1) t^{3p}_0}
\right]^2.
\end{displaymath}
Furthermore, we shall consider an action
that describes a free scalar field minimally
coupled to gravity
\begin{equation}\label{action}
I=-\int d^{4}x \  d\psi\,\sqrt{\left|\frac{^{(5)}
 g}{^{(5)} g_0}\right|} \ \left[
\frac{^{(5)} R}{16\pi G}+ ^{(5)}{\cal 
L}(\varphi,\varphi_{,A})\right],
\end{equation}
with a Lagrangian
\begin{equation}
L = \sqrt{\frac{^{(5)}g}{^{(5)}g_0}}
{\cal L}(\varphi,\varphi_{,A}),
\end{equation}
and the Lagrangian density
\begin{equation}\label{lag}
{\cal L} = \frac{1}{2} g^{AB} \varphi_{,A} \varphi_{,B}.
\end{equation}
The scalar $^{(0)} g_0$ in the action (\ref{action}) is given
by $\left.^{(0)} g\right|_{t=t_0,\psi=\psi_0}$, such that
\begin{displaymath}
\left|^{(5)}g_0\right| = \left(\frac{t_0 \psi^{\frac{4p-1}{p-1}}_0}{
(p-1)}\right)^2,
\end{displaymath}
where $t_0$ and $\psi_0$ are constants.
Note that $\left| ^{(5)} g\right|$ and
$\left| ^{(5)} g_0\right|$ are not well defined for $p=1$.
The Lagrange equation is given by
\begin{equation}
\ddot\varphi + \frac{3p+1}{t} \dot\varphi - \left(\frac{t^p_0
\psi^{\frac{1}{1-p}}}{t^p}\right)^2  \nabla^2\varphi -
\psi \frac{(p-1)(4p-1)}{t^2} \varphi_{,\psi} -
\psi^2 \frac{(p-1)^2}{t^2} \varphi_{,\psi\psi} =0,
\end{equation}
where the overdot denotes the derivative with respect to the time
and $\varphi_{,\psi} = {\partial\varphi \over \partial \psi}$.
In order to simplify the structure of this equation we propose
the transformation $\varphi = \left({t_0\over t}\right)^{{3p+1\over2 }}
\left({\psi_0 \over \psi}\right)^{{4p-1 \over 2(p-1)}} \chi$, such
that the equation of motion for $\chi$ is
\begin{equation}
\ddot\chi - \psi^{\frac{2}{1-p}} \left(\frac{t_0}{t}\right)^{2p}
\nabla^2\chi - \psi^2 \frac{(p-1)^2 }{t^2} \chi_{,\psi\psi} +
\frac{(31p^2-14p+2)}{4t^2}\chi=0.
\end{equation}
We propose the following Fourier's expansion for $\chi$
\begin{equation} \label{four}
\chi(t,\vec r,\psi) = \frac{1}{(2\pi)^{3/2}} {\Large\int} d^3 k_r
{\Large\int} dk_{\psi} \left[a_{k_r k_{\psi}}
e^{i(\vec k_r.\vec r+\vec k_{\psi}.\psi)} \xi_{k_rk_{\psi}} +
a^{\dagger}_{k_r k_{\psi}}
e^{-i(\vec k_r.\vec r+\vec k_{\psi}.\psi)}
\xi^*_{k_r k_{\psi}}\right],
\end{equation}
where the operators $a_{k_r k_{\psi}}$ and $a^{\dagger}_{k_r k_{\psi}}$
describe the algebra
\begin{displaymath}
\left[a_{k_rk_{\psi}}, a^{\dagger}_{k'_{r}k'_{\psi}}\right] =
\delta^{(3)}\left(\vec k_r - \vec k'_r \right) \delta\left(
\vec k_{\psi} - \vec k'_{\psi}\right), \quad
\left[a_{k_rk_{\psi}}, a_{k'_{r}k'_{\psi}}\right] =
\left[a^{\dagger}_{k_rk_{\psi}}, a^{\dagger}_{k'_{r}k'_{\psi}}\right] =0.
\end{displaymath}
The dymanics of the time dependent modes $\xi_{k_rk_{\psi}}$ is
given by
\begin{equation}\label{modes}
\ddot\xi_{k_rk_{\psi}} + \left[\psi^{\frac{2}{1-p}} \left(\frac{t_0}{t}
\right)^{2p} k^2_r  +
\frac{\psi^2 (p-1)^2}{t^2} \left( k^2_{\psi} +
\frac{(31p^2-14p+2)}{4(p-1)^2 \psi^2} - 2 i k_{\psi} \frac{\partial}{\partial
\psi} - \frac{\partial^2}{\partial\psi^2} \right) \right]\xi_{k_rk_{\psi}}=0.
\end{equation}
The commutator between $\chi$ and $\dot\chi$ is
\begin{equation}
\left[\chi(t,\vec r, \psi), \dot\chi(t,\vec r', \psi')\right]=
i \delta^{(3)} (\vec r - \vec r') \delta (\psi - \psi'),
\end{equation}
which complies for $\xi_{k_rk_{\psi}}(t,\psi)
\dot\xi^*_{k_rk_{\psi}}(t,\psi') - \xi^*_{k_rk_{\psi}}(t,\psi')
\dot\xi_{k_rk_{\psi}}(t,\psi) =i$, that guarantizes the normalization
of $\xi_{k_rk_{\psi}}$.
Furthermore, if we make the transformation
$\xi_{k_rk_{\psi}} = e^{-i \vec k_{\psi}. \vec\psi}
\tilde\xi_{k_rk_{\psi}}$, we obtain
from eq. (\ref{modes}) the equation of motion for $\tilde\xi_{k_rk_{\psi}}$
\begin{equation}\label{modes1}
\ddot{\tilde\xi}_{k_rk_{\psi}} -
\frac{\psi^2}{t^2} (p-1)^2 \frac{\partial^2 \tilde\xi_{k_rk_{\psi}}}{
\partial\psi^2} +
\left[\psi^{\frac{2}{1-p}} \left(\frac{t_0}{t}\right)^{2p} k^2_r +
\frac{(31 p^2 - 14 p+1)}{4 t^2} \right] \tilde\xi_{k_rk_{\psi}}=0,
\end{equation}
and the expansion (\ref{four}) for $\chi$ can be rewritten as
\begin{equation}\label{20}
\chi(t,\vec r,\psi) = \frac{1}{(2\pi)^{3/2}} {\Large\int} d^3 k_r
{\Large\int} dk_{\psi} \left[a_{k_r k_{\psi}}
e^{i\vec k_r.\vec r} \tilde\xi_{k_rk_{\psi}} +
a^{\dagger}_{k_r k_{\psi}}
e^{-i\vec k_r.\vec r}
\tilde\xi^*_{k_r k_{\psi}}\right].
\end{equation}

\subsection{4D dynamics of the inflaton field}

Now we consider a foliation (choice of a hypersurface)
$\psi=\psi_0$ on the metric (\ref{pdl}).
On this foliation, the effective 4D line element is
\begin{equation}\label{4dl}
dS^2 \rightarrow ds^2 = \psi^2_0 dt^2 - \left(\frac{t}{t_0}\right)^{2p}
\psi^{\frac{2p}{p-1}}_0 dr^2,
\end{equation}
where $\psi_0$ is a dimensionless constant and the scale factor of the
universe is given by $a = a_0 \left({t\over t_0}\right)^p$. During inflation
$p \gg 1$, such that $\ddot a >0$ and, for $\dot\beta = H = \dot a/a$
in eq. (\ref{9a}), the equation of state describes a quasi de Sitter (quasi
vacuum) expansion: ${\rm p} \simeq \left( {4\over 3} {\dot H\over H^2}
+1\right) \rho_t \simeq - \rho_t$.
The
Lagrangian density (\ref{lag}) can be expanded in the following
manner: $^{(5)} {\cal L}
= {1\over 2} \left[ g^{\mu\nu} \varphi_{,\mu}\varphi_{,\nu}
+ g^{\psi\psi} \varphi_{,\psi}\varphi_{,\psi}\right]$,
such that the effective 4D Lagrangian density
for the inflaton field on the foliation $\psi=\psi_0$
can be written as
\begin{equation}
^{(4)} {\cal L} =  \left.
\frac{1}{2} g^{\mu\nu} \varphi_{,\mu} \varphi_{,\nu} - V(\varphi)
\right|_{\psi=\psi_0},
\end{equation}
where $V(\varphi)$ is given by
\begin{equation}
V(\varphi) = -\left.\frac{1}{2} g^{\psi\psi} \varphi_{,\psi}
\varphi_{,\psi}\right|_{\psi=\psi_0}
=\left.
\frac{(p-1)^2}{2t^2} \left(\frac{\partial\varphi}{\partial\psi}\right)^2
\right|_{\psi=\psi_0}.
\end{equation}
On the other hand, for a foliation $\psi=\psi_0$, the dynamics of the
inflaton field holds
\begin{equation}\label{25}
\ddot\varphi + \frac{(3p+1)}{t} \dot\varphi -
\left(\frac{t_0}{t}\right)^{2p} \psi^{\frac{2}{1-p}} \nabla^2\varphi
-\left.\left[\frac{\psi(p-1)(4p-1)}{t^2} \varphi_{,\psi} +
\frac{\psi^2(p-1)^2}{t^2} \varphi_{,\psi\psi}\right]\right|_{\psi=\psi_0}=0,
\end{equation}
so that we can make the following identification:
\begin{equation}
V'(\varphi) =\left.\left[ \frac{\dot\varphi}{t}
-\frac{\psi(p-1)(4p-1)}{t^2} \varphi_{,\psi} -
\frac{\psi^2(p-1)^2}{t^2} \varphi_{,\psi\psi}\right]\right|_{\psi=\psi_0}.
\end{equation}

Furthermore, the scalar field $\chi$ in eq. (\ref{20}) on the foliation
$\psi=\psi_0$ now holds
\begin{equation}\label{ul}
\chi(t,\vec r, \psi=\psi_0) = \frac{1}{(2\pi)^{3/2}}
{\Large\int} d^3 k_r {\Large\int} dk_{\psi} \left[ a_{k_r k_{\psi}}
e^{i \vec k_r . \vec r} \tilde\xi_{k_r k_{\psi}} + c.c. \right]
\delta(k_{\psi } - k_{\psi_0}).
\end{equation}
Note that the equation (\ref{25}) is not separable on the coordinate
$\psi$. However, in the limit $p \gg 1 $, which is relevant to study
the inflationary expansion of the early universe, the term
$\left({t_0\over t}\right)^{2p} \psi^{{2\over 1-p}} \nabla^2\varphi$ in eq.
(\ref{25}) tends asymptotically to
$\left({t_0\over t}\right)^{2p} \nabla^2\varphi$ as $p \rightarrow \infty$.
Hence, in this limital case the dynamics for $\varphi$ is governed
by the equation
\begin{equation}
\ddot\varphi + \frac{(3p+1)}{t} \dot\varphi -
\left(\frac{t_0}{t}\right)^{2p}  \nabla^2\varphi
-\left.\left[\frac{\psi(p-1)(4p-1)}{t^2} \varphi_{,\psi} +
\frac{\psi^2(p-1)^2}{t^2} \varphi_{,\psi\psi}\right]\right|_{\psi=\psi_0}=0,
\end{equation}
which is separable on the variable $\psi$ and more easily workable.
This particular case will be studied in the next section.

\section{Inflationary expansion ($p\gg 1$)}

We consider the case where $p \gg 1$, which
describes an inflationary power-law expansion in the very
early universe.
In this case the equation for the modes (\ref{modes1}), can be
approximated to
\begin{equation}\label{modes2}
\ddot{\tilde\xi}_{k_rk_{\psi}} -  \frac{\psi^2 (p-1)^2}{t^2}
\frac{\partial^2  \tilde\xi_{k_rk_{\psi}}}{\partial\psi^2} +
\left[ k^2_r \left(\frac{t_0}{t}\right)^{2p}
+ \frac{(31 p^2 - 14 p +2)}{4 t^2} \right] \tilde\xi_{k_rk_{\psi}} \simeq 0.
\end{equation}
The general solution for this equation on the foliation $\psi=\psi_0$
is
\begin{equation}
\tilde\xi_{k_rk_{\psi}}(t,\psi) = \alpha \sqrt{\frac{t}{t_0}}
{\cal H}^{(1)}_{\nu}[x(t)],
\end{equation}
where $\nu = \sqrt{{1+4C_1 \over 2(p-1)}}$, $x(t) = k_r t^p_0 t^{1-p}/(p-1)$,
$\alpha = A_1
\left[ M_1
\psi^{{1\over 2}+\sqrt{{1\over 4}+{C \over (p-1)^2}}}_0
+ M_2 \psi^{{1\over 2}-\sqrt{{1\over 4}+{C \over (p-1)^2}}}_0\right]$
$C = C_1 + {(31 p^2 - 14 p +2)\over 4}$ and $C_1$, $A_1$, $M_1$ and
$M_2$ are constants of
integration.
Furthermore, the normalization condition implies that
\begin{equation}
\frac{4(p-1)}{t_0\pi} \left|\alpha\right|^2=1.
\end{equation}
To calculate the inflaton field fluctuations on the infrared sector, which
is relevant for super Hubble scales during inflation, we can make
use of the asymptotic representations of the first kind Hankel function
${\cal H}^{(1)}_{\nu}[x]$ for $x \ll 1$
\begin{displaymath}
{\cal H}^{(1)}_{\nu}[x] \simeq \frac{1}{\Gamma(\nu+1)} \left(
\frac{x}{2}\right)^{\nu} + \frac{i}{\pi} \Gamma(\nu)
\left(\frac{x}{2}\right)^{-\nu}.
\end{displaymath}
With this representation the squared $\chi$ fluctuations
on cosmological scales are
$\left.\left<\chi^2\right>\right|_{IR}
= {1\over 2\pi^2} {\Large\int}^{\epsilon k_0}_0 dk_r k^2_r
\tilde\xi_{k_rk_{\psi_0}} \tilde\xi^*_{k_rk_{\psi_0}}$,
where $k_0(t) = {\sqrt{31 p^2 -14 p +2} \over 2 t^p_0} t^{p-1}$
is the wavenumber that separates the infrared (IR) and ultraviolet
sectors.
Making the calculation, we obtain
\begin{eqnarray}
\left.\left<\chi^2\right>\right|_{IR} & = & \frac{t}{8\pi (p-1)}
\left[\frac{1}{\Gamma^2(\nu+1)} \left(\frac{\epsilon^3(31p^2 -
14 p+2) t^{2(p-1)}}{16 (p-1) t^{2p}_0}\right)^{2\nu}\frac{1}{(2\nu+3)}
\right.      \nonumber \\
& - & \left.\frac{\Gamma^2(\nu)}{\pi^2 (3-2\nu)} \left(
\frac{16 (p-1) t^{2p}_0 t^{-2(p-1)}}{\epsilon^3(31 p^2 -14 p +2)}
\right)^{2\nu}\right],
\end{eqnarray}
and the squared $\varphi$ fluctuations, which are the relevant for us, are
\begin{eqnarray}
\left.\left<\varphi^2\right>\right|_{IR} & = & \frac{t^{-3p}
t^{(3p+1)}_0}{8\pi (p-1)}
\left[\frac{1}{\Gamma^2(\nu+1)} \left(\frac{\epsilon^3(31p^2 -
14 p+2) t^{2(p-1)}}{16 (p-1) t^{2p}_0}\right)^{2\nu}\frac{1}{(2\nu+3)}
\right. \nonumber \\
& -& \left. \frac{\Gamma^2(\nu)}{\pi^2 (3-2\nu)} \left(
\frac{16 (p-1) t^{2p}_0 t^{-2(p-1)}}{\epsilon^3(31 p^2 -14 p +2)}
\right)^{2\nu}\right],
\end{eqnarray}
which decrease with time as
$ \left.\left< \varphi^2\right>\right|_{IR}\sim t^{-3p}$.
The power spectrum of these fluctuations goes as ${\cal P}(k_r)
\sim k^{3-2\nu}_r$ which is scale invariant for $C_1 = {9(p-1)-2\over 8} $
and hence for $C = {61 p^2 -(19 p + 7) \over 8}$.
This result is very interesting because (with an
adequate choice of constant values) we can obtain scale invariance
for the power spectrum of $\left<\varphi^2\right>$ for any $p \gg 1$,
independently of some particular value for it.

\section{Final comments}

In this paper we have studied power-law inflation in the
limital case $p \gg 1$ from a STM theory of gravity using
the Ponce de Leon metric. In this approach, the inflationary
expansion is governed by a single scalar field, that, on a
foliation $\psi=\psi_0$ in the 5D metric (\ref{pdl}) can be
identified as the inflaton field evolving on the effective 4D
Friedmann-Robertson-Walker metric (\ref{4dl}).
In the Wesson's theory [called Space-Time-Matter (STM) theory],
the extra dimension is not assumed to be compactified, which is a major
departure from earlier multidimensional theories where the cylindricity
conditions was imposed. In this theory, the original motivation for assuming
the existence of a large extra dimension was to achieve the unification of
matter and geometry, i.e. to obtain the properties of matter as a consequence
of the extra dimensions.
A very important fact in our approach
is that the effective potential $V(\varphi)
=-\left.{1\over 2} g^{\psi\psi}
\left({\partial \varphi\over\partial \psi}\right)^2\right|_{\psi=\psi_0}$,
has a geometrical origin.

In this paper we have studied with major detail
the case $p \gg 1$, because
is the only where the field $\varphi(t, \vec r, t)$ is
separable on the variable $\psi$ and can be more easily treated.
However, for cases where the power $p$ is of the order of the unity, the
fifth coordinate $\psi$ could play an more important role in the
spectrum of the squared $\varphi$ fluctuations.\\
\vskip .3cm
\noindent
\centerline{{\bf Acknowledgements}}
\vskip .1cm
\noindent
MA and MB acknowledge CONICET (Argentina) and UNMdP for financial
support.\\

\end{document}